\documentstyle[preprint,eqsecnum,aps]{revtex}

\begin{document}

\draft

\title{Comments on the paper astro-ph/0103335 by C Rubano and P Scudellaro}

\author{Rolando Cardenas\thanks{rcardenas@mfc.uclv.edu.cu}, Tame Gonzalez\thanks{tame@mfc.uclv.edu.cu}, Osmel Martin\thanks{osmel@mfc.uclv.edu.cu}, Israel Quiros\thanks{israel@mfc.uclv.edu.cu} and Diosdado Villegas\thanks{villegas@mfc.uclv.edu.cu}}
\address{Departamento de Fisica. Universidad Central de Las Villas. Santa Clara. CP: 54830 Villa Clara. Cuba}
 
\date{\today}

\maketitle

\begin{abstract}
We briefly comment on a paper by Rubano and Scudellaro [astro-ph/0103335] where they found general exact solutions for two classes of exponential potentials in a scalar field model for quintessence. In that paper the authors were led to some interesting conclusions after a proper choice of the integration constants. By using dimensionless variables we show that the integration constants can be found explicitly without additional assumptions. In consequence we revise some results and conclusions in that paper. We also reproduce observations for Type-Ia supernovae with good accuracy. 
\end{abstract}


\section{Introduction}

In a recent paper\cite{rubano-scudellaro} Rubano and Scudellaro found general exact solutions for two classes of exponential potentials in a scalar field model for quintessence. In that paper the authors studied a two-component perfect fluid (dust plus a scalar field) and considered an exponential potential and a combination of two exponential potentials for the quintessence field respectively. The quintessense action is given by $S=\int d^4 x\;\sqrt{-g}\{\frac{c^2}{16\pi G}R+{\cal L}_\phi+{\cal L}_{m}\}$,
where ${\cal L}_{m}$ is the Lagrangian for the matter degrees of freedom and the Lagrangian for the quintessense field is given by ${\cal L}_{\phi}=-\frac{1}{2}\phi_{,n} \phi^{,n}-V(\phi)$. Two classes of exponential potentials are of interest:

\begin{equation}
V_1(\phi)=B^2 e^{-\sigma \phi},
\end{equation}
and,

\begin{equation}
V_2(\phi)=A^2 e^{\sigma \phi}+B^2 e^{-\sigma \phi},
\end{equation}
where $\sigma^2=\frac{12\pi G}{c^2}$ and $A^2$ and $B^2$ are generic constants.

In Ref.\cite{rubano-scudellaro} a flat (k=0) FRW universe was studied and the corresponding field equations that are derivable from an action principle are the following,

\begin{equation}
(\frac{\dot{a}}{a})^2=\frac{2}{9}\sigma^2 \{ \frac{D}{a^3}+\frac{1}{2} \dot{\phi}^2 + V_j(\phi) \},
\end{equation}

\begin{equation}
2\frac{\ddot{a}}{a}+(\frac{\dot{a}}{a})^2=-\frac{2}{3}\sigma^2 \{ \frac{1}{2}\dot{\phi}^2- V_j(\phi)\},
\end{equation}
and

\begin{equation}
\ddot{\phi}+3\frac{\dot{a}}{a}\dot{\phi}+V_j'(\phi)=0,
\end{equation}
where $D$ is the amount of matter (see Ref.\cite{rubano-scudellaro}), the dot means derivative in respet to the cosmic time, the comma denotes derivative in respect to $\phi$, and $j=1,2$. In that paper Rubano and Scudellaro were able to find general exact solutions to the above system of equations by introducing a pair of new variables:

\begin{equation}
a^3=u v,\;\; \phi=-\frac{1}{\sigma}\ln(\frac{u}{v}),
\end{equation}
in the first case (potential $V_1$, Eq. (1)), and

\begin{equation}
a^3=\frac{u^2-v^2}{4},\;\;\phi=\frac{1}{\sigma}\ln[\frac{B(u+v)}{A(u-v)}],
\end{equation}
in the second case (potential $V_2$, Eq. (2)). The general solutions they found are:
\begin{equation}
u(t)=u_1 t+ u_2,\;\;v(t)=\frac{\sigma^2 B^2}{6}u_1 t^3+\frac{\sigma^2 B^2}{2}u_2 t^2+v_1 t+v_2, 
\end{equation}
for the first class of potential $V_1(\phi)$ and,
\begin{equation}
u(t)=\alpha e^{\omega t}+\beta e^{-\omega t},\;\;v(t)=v_1\sin{(\omega t+v_2)}=\hat v_1\sin{(\omega t)}+\hat v_2\cos{(\omega t)}, 
\end{equation}
for a combination of two exponentials $V_2(\phi)$, where $\omega=AB\sigma^2$.

In Ref.\cite{rubano-scudellaro} the authors studied different situations by properly choosing the integration constants in Eqs. (1.8) and (1.9). The point of the present paper is to show that the integration constants in Eqs. (1.8) and (1.9) can be explicitly found without making additional assumptions if we introduce dimensionless variables. In effect, let us introduce the dimensionless time variable $\tau=H_{0} t$, where $t$ is the cosmic time and $H_{0}$ is the present value of the Hubble parameter, and the dimensionless scale factor $a(\tau)=\frac{a(t)}{a(0)}$. In these variables $H(\tau)={\dot a(\tau)}/a(\tau)$ where, from now on, the dot means derivative in respect to the dimensionless time $\tau$. Then we have that, at present $(\tau =0)$,

\begin{equation}
a(0)=1,\;\;\dot{a}(0)=1\Rightarrow H(0)=1.
\end{equation}

Besides the changes $t\rightarrow\frac{\tau}{H_0}$ and $a(t)\rightarrow a(0)\;a(\tau)$ in equations (1)-(5), one should replace $\frac{A^2}{H_{0}^{2}}\rightarrow A^2$, $\frac{B^2}{H_{0}^{2}}\rightarrow B^2$, $\frac{D}{a_{0}^{3}H_{0}^{2}}=\frac{\rho_{m_{0}}}{H_{0}^{2}}\rightarrow D$, where $\rho_{m_{0}}$ is the density of matter. After this rescaling, one can check that $\sigma^2 D=\frac{9}{2}\Omega_{m_0}$ and, besides, Friedmann Eq. (1.3) evaluated at $\tau=0$ implies for quintessence $\Omega_{Q_0}=1-\Omega_{m_0}$. The net result is that equations (1.1)-(1.5) are unchanged but the constants and parameters, for instance $\omega=AB\sigma^2$, are now dimensionless. While finding the four integration constants in Eqs. (1.8) and (1.9) respectively, we can use the two conditions in Eq. (1.10) and two of the field equations (1.3)-(1.5) (for instance (1.3) and (1.4)) evaluated at $\tau=0$. 

\section{An exponential potential}

For the potential of the class $V_1$ (Eq. (1.1)), the solutions are
\begin{equation}
u(\tau)=u_1 \tau+ u_2,\;\;v(\tau)=\frac{\omega}{6}u_1 \tau^3+\frac{\omega}{2}u_2 \tau^2+v_1 \tau+v_2, 
\end{equation}

The integration constants are the following

\begin{equation}
u_2^{(\pm)}=\pm\frac{\sqrt{3}}{2}\sqrt{\frac{2(2-q_0)-3\Omega_{m_0}}{\sigma^2 B^2}},
\end{equation}
\begin{equation}
v_2^{(\pm)}=\frac{1}{u_2^{(\pm)}},\;\;\;u_{1\;[\pm]}^{(\pm)}=\frac{3+[\pm]\frac{\sqrt{3}}{2}\sqrt{2(1+q_0)-3\Omega_{m_0}}}{2}u_2^{(\pm)},
\end{equation}
and,

\begin{equation}
v_{1\;[\pm]}^{(\pm)}=\frac{3-[\pm]\frac{\sqrt{3}}{2}\sqrt{2(1+q_0)-3\Omega_{m_0}}}{2u_2^{(\pm)}},
\end{equation}
where $q_0=-(1+\dot H(0))$ is the present value of the deceleration parameter and, the $(\pm)$ and $[\pm]$, allow, in principle, for four different branches of the solution (2.1). 

The relation between the cosmological time $t$ of \cite{rubano-scudellaro} and our conformal time $\tau$ is 
\begin{equation}
t=\frac{1}{H_0}(\tau+\frac{u_2}{u_1}).
\end{equation}
Substituting the above result in equation (1.8) and comparing with our solution gives the relation between our more general integration constants and those (denoted with a bar here) found in the particular case of \cite{rubano-scudellaro}:
\begin{equation}
\bar u_1=H_0u_1, \bar u_2=0, \bar v_1=H_0(v_1-\frac{\omega u_2^2}{2u_1}), 
\bar v_2=v_2-\frac{\omega u_2^2}{6u_1}, \bar \omega=\omega H_0^2
\end{equation}

Due to signs combination, indeed there are only two different solutions for the scale factor:
\begin{equation}
a^3(\tau)=\frac{(3\pm R_1)^2 R_2^2}{24}\tau^4+\frac{(3\pm R_1)R_2^2}{3}\tau^3+\frac{2 R_2^2-R_1^2+9}{4}\tau^2+3\tau+1,
\end{equation}
where

\begin{equation}
R_1=\frac{\sqrt{3}\sqrt{2(1+q_0)-3\Omega_{m_0})}}{2},\;\; R_2=\frac{\sqrt{3}\sqrt{2(1-q_0)-3\Omega_{m_0})}}{2},
\end{equation}
For $R_1$ to be real, in Eq. (2.8), $\sqrt{2(1+q_0)-3\Omega_{m_0}}$ should be real, then the following constraint holds,

\begin{equation}
q_0\geq -1+\frac{3}{2}\Omega_{m_0}.
\end{equation}

In the limiting situation in which $q_0= -1+\frac{3}{2}\Omega_{m_0}$,\footnotemark\footnotetext{This condition can be written, also, in the equivalent form $q_0=\frac{\Omega_{m_0}}{2}-\Omega_{Q_0}$, where the quintessence field stands for a dynamical cosmological constant. This relationship is often used in the bibliography on accelerated expansion\cite{binetruy,data}} $R_1=0$, which implies that survive just one branch of the solution.

In general, from Eqs. (2.8) and (2.7), one sees that the relevant magnitudes characterising the evolution of the universe depend only on two parameters $q_0$ and $\Omega_{m_0}$. These are not sensible to the particular value the dimensionless constant $\omega=\sigma^2 B^2$ takes.

 Now we check the singular points of the  solutions found. The condition $a(\tau)=0$ means $u(\tau)=0$ and/or $v(\tau)=0$, i. e., we have four roots in $\tau$. The one derived from the condition $u(\tau)=0$ is

\begin{equation}
\tau_{in}=-\frac{2}{3+[\pm]\frac{\sqrt{3}}{2}\sqrt{2(1+q_0)-3\Omega_{m_0}}}.
\end{equation}

If $\tau_{in}$ were positive then we were faced with a very serious problem since $\tau=0$ at present. However a careful analysis of (2.10) shows that, for the positive branch, $\tau_{in}$ is always negative, meanwhile for the negative branch, it could be positive only if the following condition holds: $q_0>5+3\Omega_{m_0}/2$. But this condition is incompatible with present accelerated expansion. In case $v(\tau)=0$ we found three roots, two of which are imaginary and one real. The real root is very complicated and we will not to write it here. A careful analysis shows that this root for $\tau_{in}$ is negative as well, besides, by its absolute magnitude it is always greater than $\tau_{in}$ in Eq. (2.10) (i. e., it is prior to $\tau_{in}$ in Eq. (2.10)). Besides during the period between both roots the scale factor becomes imaginary so we drop times prior to $\tau_{in}$ in (2.10). This is evident from Figs. 1 and 2, where the behavior of the scale factor is shown for both branches of the solution for the particular values $\Omega_{m_0}=0.3$ and $q_0=-0.4$ and $q_0=-0.45$ respectively. Similar arguments hold for the double exponential potential. The behavior of the scale factor of the universe shows that $a(\tau)$ has zeroes for positive $\tau$ in no case.

Now we proceed to analyze how our solution reproduces experimental results. With this purpose, in Fig. 3 we plot the distance modulus $\delta(z)$ vs redshift z, calculated by us and the one obtained with the usual model with a constant $\Lambda$ term. The relative deviations are of about 0.5$\%$.   

\section{A more complicated potential}

For the second potential $V_2(\phi)$ (Eq. (1.2)) we write solutions in the following way:
\begin{equation}
u(\tau)=\alpha e^{\omega \tau}+\beta e^{-\omega \tau},\;\;v(\tau)=v_1\sin{(\omega \tau)}+v_2\cos{(\omega \tau)}, 
\end{equation}

Here the situation is more complex. There are 8 branches of the solution. We studied one of these branches, leaving for further investigation a detailed study of the remaining ones. In the case studied here the integration constants $\alpha$, $\beta$, $v_1$, and $v_2$ in the solution (3.1) are found to be

\begin{eqnarray}
v_1&=&\frac{\sqrt{4+v_2^2}\sqrt{2(9-9\Omega_{m_0}-3(1-2q_0))}+6v_2}{4\omega},\;\;
v_2=\sqrt{\frac{3(1-2q_0)+9-9\Omega_{m_0}}{4\omega}},\nonumber\\
&\alpha&=\sqrt{4+v_2^2}-\beta,\;\;
\beta=\frac{\omega(4+v_2^2)-6-\omega v_1v_2}{2\omega\sqrt{4+v_2^2}}.
\end{eqnarray}

As before the situation simplifies if we consider the limiting case when $q_0=-1+\frac{3}{2}\Omega_{m_0}$. In this case the above integration constants look like

\begin{equation}
v_1=\frac{3v_2}{2\omega},
v_2=3\sqrt{\frac{1-\Omega_{m_0}}{2\omega}}.
\end{equation}
We have three parameters in the model with this potential ($A$, $B$ and $\omega$). For simplicity, we have set $\omega=1$ in the following calculations.

Figure 4 shows the behaviour of the scale factor for $\Omega_{m_0}=0.3$ and $q_0=-0.45$, while figure 5 plots $\delta(z)$ vs redshift $z$ and compares with the usual experimental model with a constant $\Lambda$ term\cite{rubano-scudellaro}. Relative deviations in this case are of about 0.25$\%$.

\section{Conclusions}

We showed that using the technique of adimensional variables and shifting the time origin, it was possible to find more general solutions than those found in Ref \cite{rubano-scudellaro}.

In most cases we have used the simplifying condition $q_0= -1+\frac{3}{2}\Omega_{m_0}$, which is justified at least by the fact that a good agreement with experimental results is achieved. Indeed, we detected very small variations of $\delta(z)$ with relatively wide variations of $\Omega_{m_0}$, always within the aforementioned condition. This led us to explore another measurable quantities like the age of the universe.
Equation (2.5) may be writen

\begin{equation}
H_0 t=\tau+\frac{u_2}{u_1}.
\end{equation}

Evaluating the above equation for the present ($\tau=0$),

\begin{equation}
H_0 t_0=\frac{u_2}{u_1},
\end{equation}
where $t_0$ is the age of the universe. Using the equations (2.2) and (2.3) and the simplifying condition $q_0= -1+\frac{3}{2}\Omega_{m_0}$ yields

\begin{equation}
H_0 t_0=\frac{2}{3},
\end{equation}
This is in agreement with the accepted standard cosmological paradigm ($\bar H_0\bar t_0$ should be of order unity). However, better agreement is obtained if we use instead the more general requirement (2.9). As an illustration, consider the solution (2.7) for the single exponetial potential. From Eq. (2.10) it is seen that, for the negative branch of the solution (2.7) $\frac{2}{3}<H_0 t_0\leq 1$ in all cases. For the positive branch $H_0 t_0<\frac{2}{3}$ and so, from experiment we see that the negative branch of the solution is more appropiated than the positive one.
 
As seen from the figures, the model for quintessence studied in Ref.\cite{rubano-scudellaro} yields an eternally accelerating universe with an event horizon that seems to be incompatible with superstring theory\cite{hwang}. One possible way to make this model compatible with observational evidence for a presently accelerating universe and with the absence of event horizons, is to add a negative constant term to the potentials $V_j$, equivalent to having a negative cosmological constant\cite{hwang}. In a forthcoming paper we explore this possibillity.

This short paper is dedicated to the memory of our colleague and friend Angelo Gino Agnese who introduced us to this subject. 

We acknowledge Andro Gonzales and Livan Rivero for help in computations and useful comments. We thank the MES of Cuba by financial support of this research.


\begin{figure}[b]
\caption{The evolution of the scale factor is shown for the two branches of the solution for the single exponential potential for $\Omega_{m_0}=0.3$ and $q_0=-0.4$.} 
\end{figure}

\begin{figure}[b]
\caption{The evolution of the scale factor is shown for the two branches of the solution for the single exponential potential for $\Omega_{m_0}=0.3$ and $q_0=-0.45$. The two roots for which $a(\tau)=0$ are visible. For the period between both roots the scale factor is imaginary, so this region is physically meaningless} 
\end{figure}

\begin{figure}[b]
\caption{The modulus distance $\delta(z)$ is ploted as a function of the redshift $z$. Dots represent the experimental values according to the standard metodology for $\Omega_{m_0}=0.3$. The solid line represents the theoretical curve obtained in the present model if we consider a single exponential potential for $\Omega_{m_0}=0.3$ and $q_0=-0.55$} 
\end{figure}

\begin{figure}[b]
\caption{The evolution of the scale factor is shown for the two branches of the solution for the double exponential potential for $\Omega_{m_0}=0.3$ and $q_0=-0.45$.} 
\end{figure}

\begin{figure}[b]
\caption{The modulus distance $\delta(z)$ is ploted as a function of the redshift $z$. Dots represent the experimental values according to the standard metodology. The solid line represents the theoretical curve obtained in the present model if we consider a single exponential potential. In both cases $\Omega_{m_0}=0.3$. We chose $q_0=-0.55$} 
\end{figure}


\begin{thebibliography}{99}


\bibitem{rubano-scudellaro}  C. Rubano and P. Scudellaro, astro-ph/0103335 (To appear Gen. Rel. Grav).

\bibitem{binetruy} P. Binetruy, Int. J. Theor. Phys. \textbf{39} (2000) 1859-1875
 (hep-ph/0005037).

\bibitem{data}  S. Perlmutter et al (Supernova Cosmology Project Collaboration), Astrophys. J. \textbf{517} (1999) 565-586 (astro-ph/9812133); Adam G. Riess et al (Supernova Search Team Collaboration), Astron. J. \textbf{116} (1998) 1009-1038 (astro-ph/9805201);  Michael S. Turner, astro-ph/9904049 (To be published in the proceedings of Type Ia Supernovae: Theory and Cosmology (held at the University of Chicago, 29 - 31 October 1998), edited by Jens Niemeyer and James Truran (Cambridge Univ. Press, Cambridge, UK); Lawrence M. Krauss and Michael S. Turner, Gen. Rel. Grav. \textbf{31} (1999) 1453-1459 (astro-ph/9904020).

\bibitem{hwang}  Je-An Gu and W-Y. P. Hwang, astro-ph/0106387.



 \end{thebibliography}
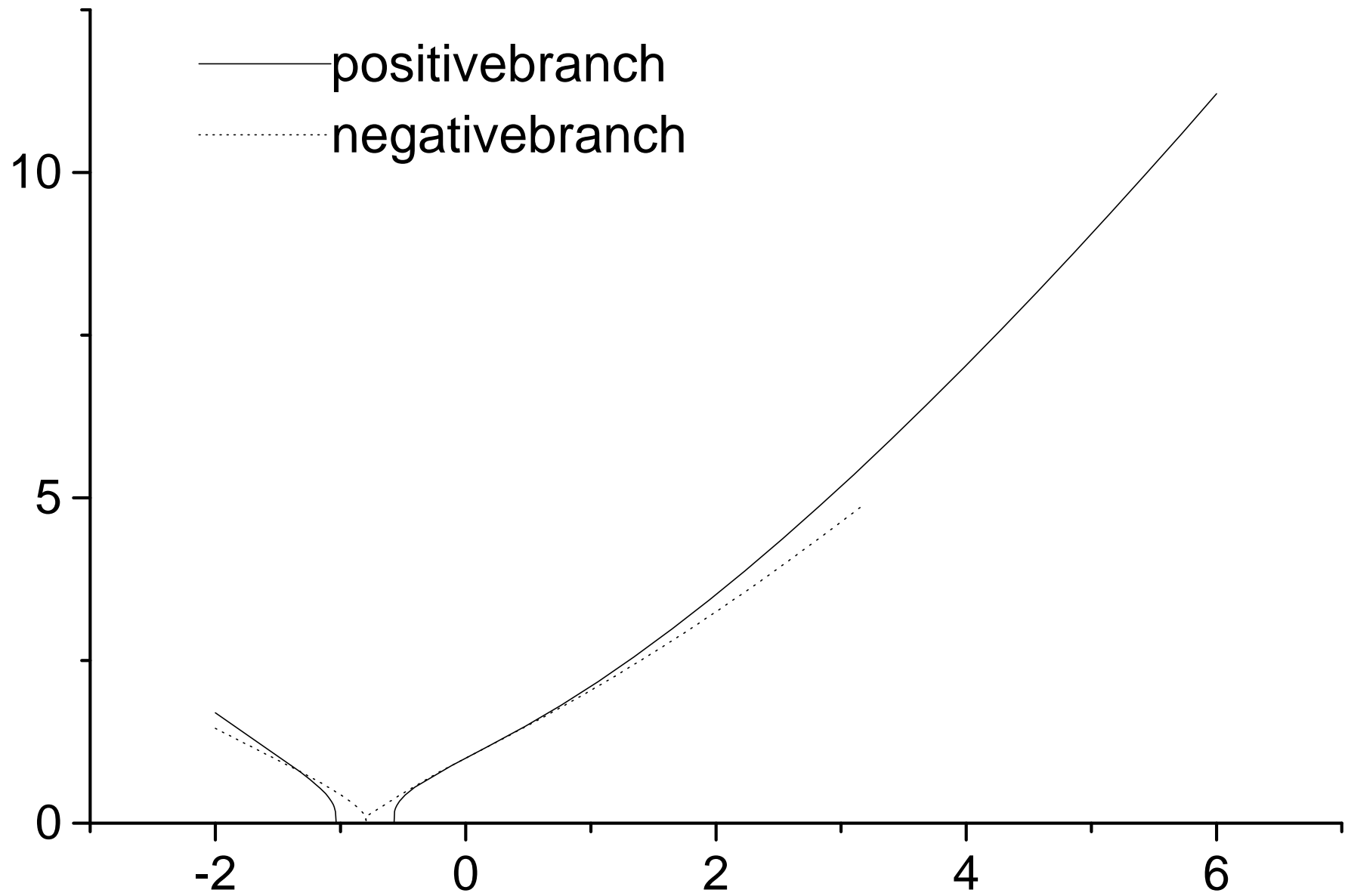
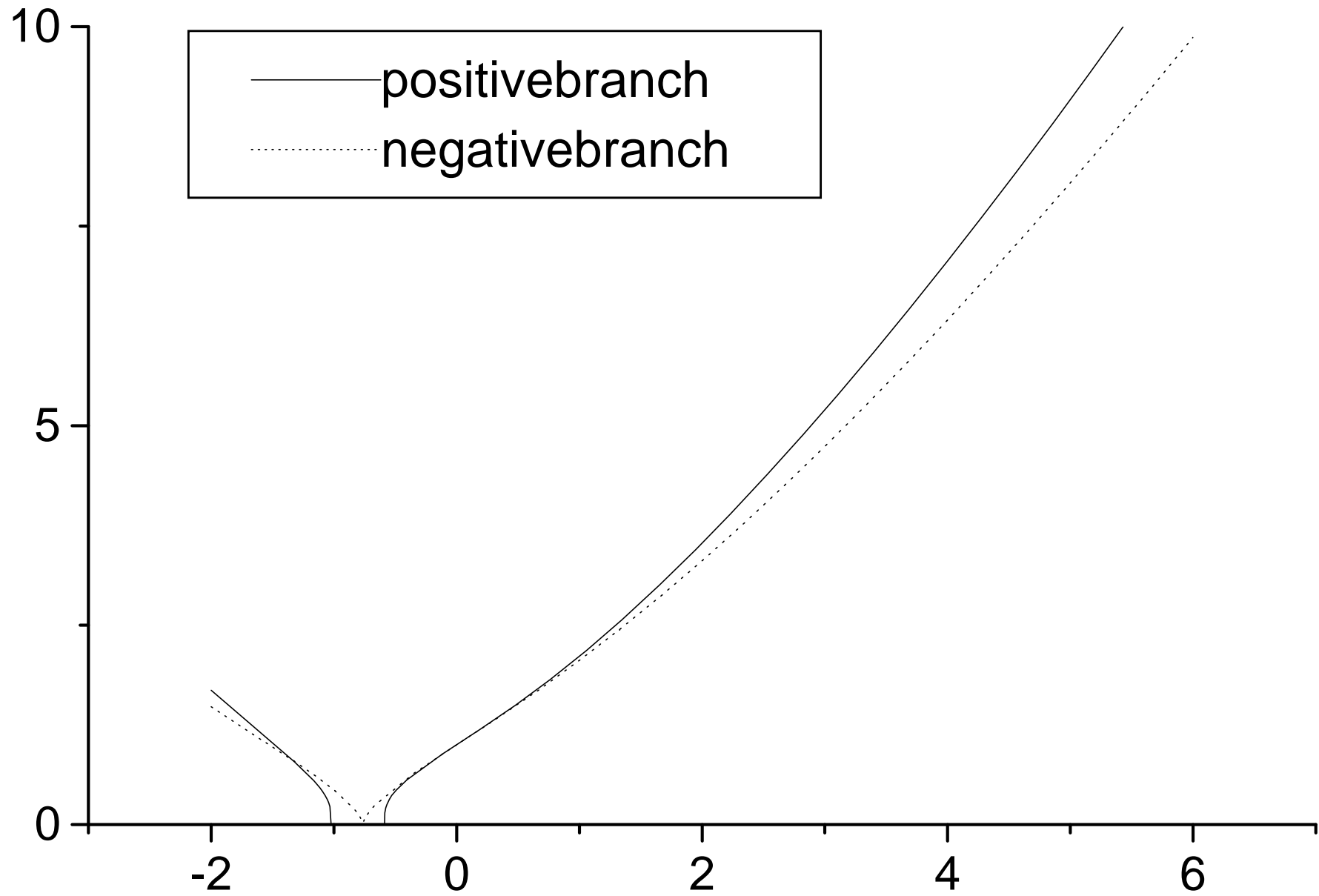
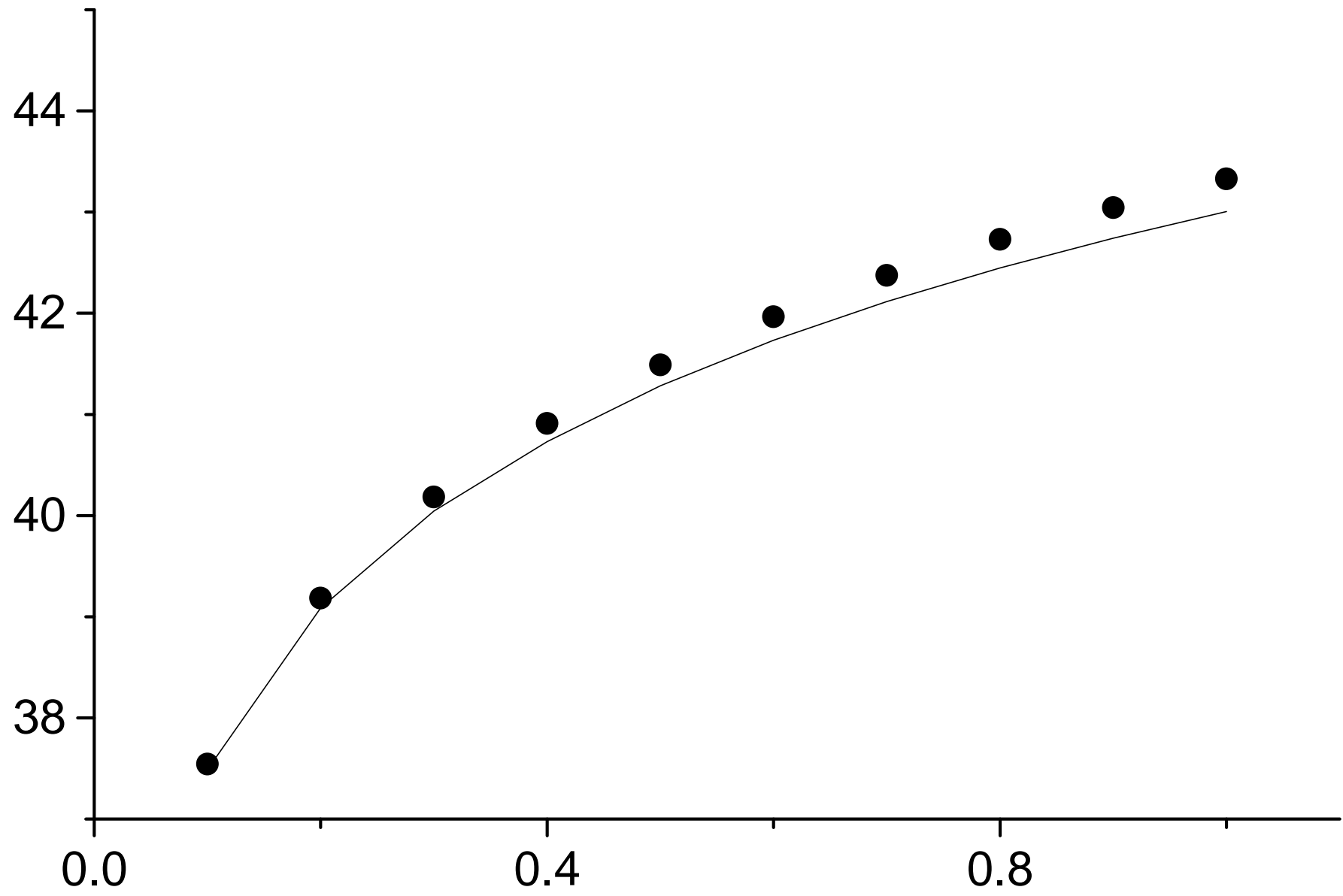
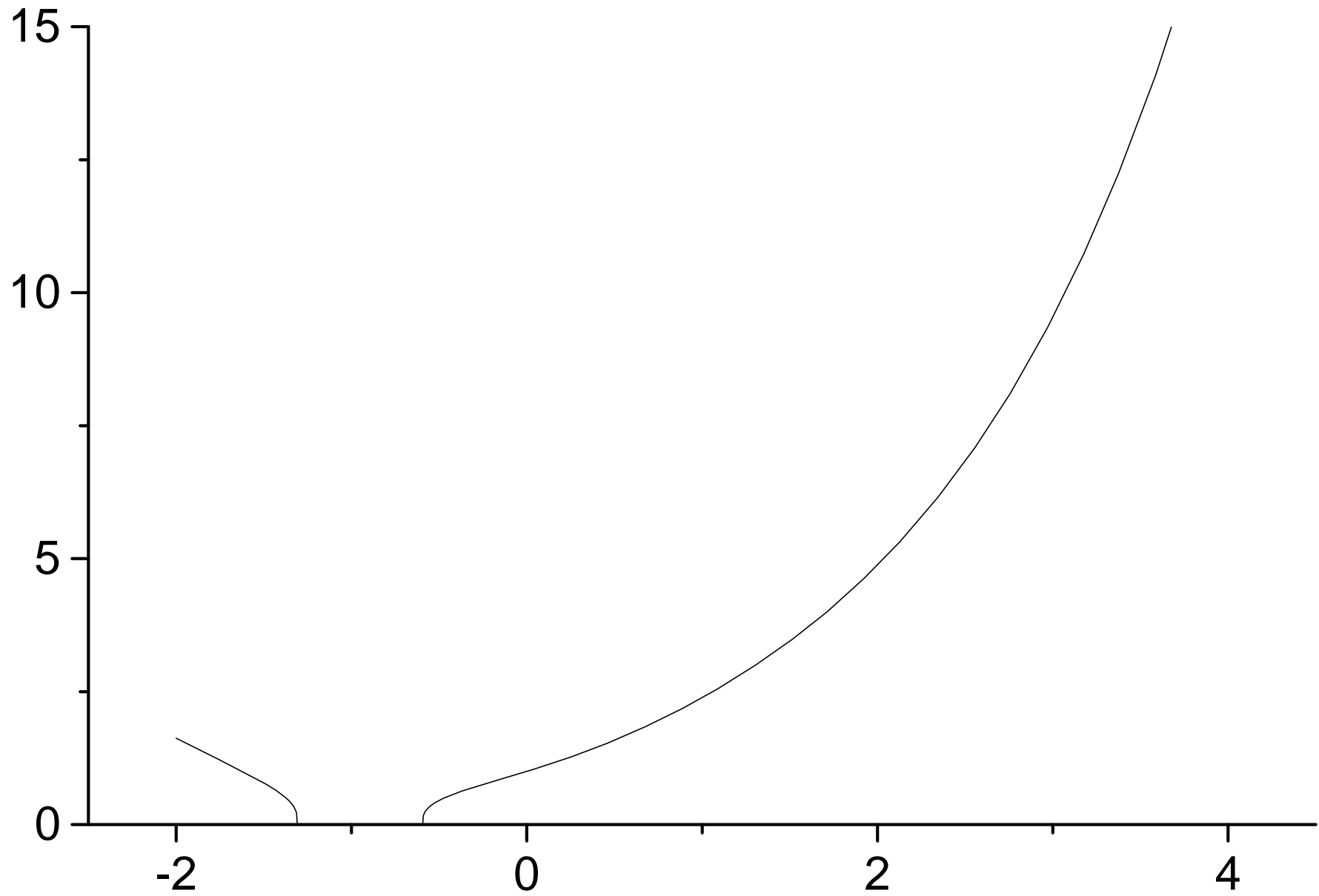
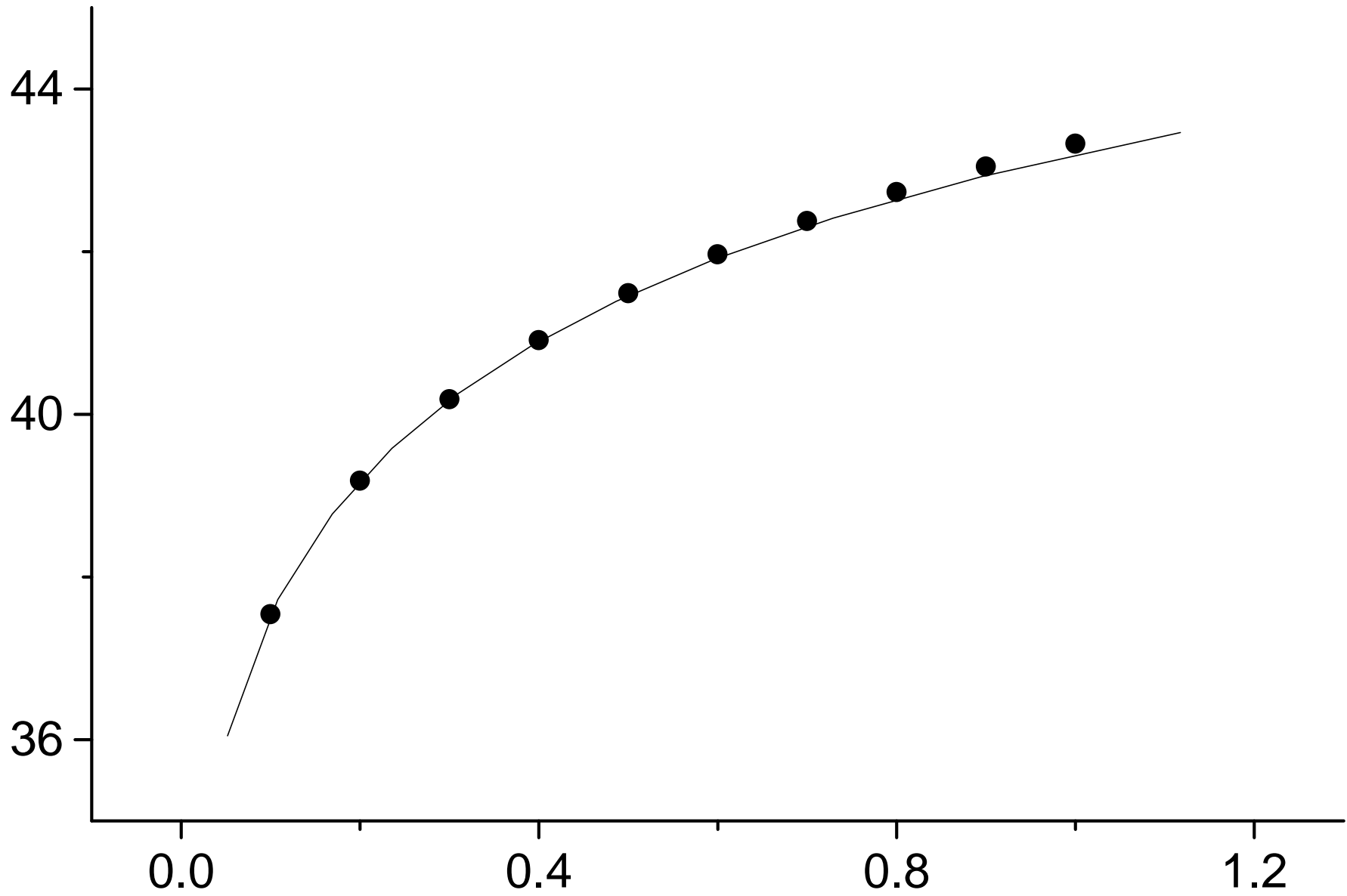
\end{document}